\documentclass[pra,onecolumn,showpacs,preprintnumbers,amsmath,amssymb]{revtex4}

\usepackage[english]{babel}
\usepackage[latin1]{inputenc}
\usepackage{amsmath,amsfonts,amssymb}
\usepackage{bm,graphicx,graphics,color}
\usepackage{amsmath}
\usepackage{amssymb}
\usepackage{graphicx}
\usepackage{graphics}
\usepackage{bm}
\usepackage{epstopdf}
\usepackage[compatibility=false]{caption}
\usepackage{subcaption}
\usepackage{color}

\begin{document}

\title {Effects of the Fermionic Vacuum Polarization in QED}

\author{M.F.X.P. Medeiros}
 \email{xavier.pmedeiros@yahoo.com.br}
\affiliation{IFQ - Universidade Federal de Itajub\'a, Av. BPS 1303, Caixa Postal 50, CEP. 37500-903, Itajub\'a, MG, Brazil.}
\author{F. E. Barone}
 \email{febarone@cbpf.br} 
\author{F. A. Barone}
 \email{fbarone@unifei.edu.br} 
\affiliation{IFQ - Universidade Federal de Itajub\'a, Av. BPS 1303, Caixa Postal 50, CEP. 37500-903, Itajub\'a, MG, Brazil.}

\date{}

\baselineskip=20pt

\begin{abstract}
Some effects of vacuum polarization in QED due to the presence of field sources are investigated. We focus on effects with no counter-part in Maxwell electrodynamics. The the Uehling interaction energy between two stationary point-like charges is calculated exactly in terms of Meijer-$G$ functions. Effects induced on a hydrogen atom by the vacuum polarization in the vicinity of a Dirac string are considered. We also calculate the interaction between two parallel Dirac strings and corrections to the energy levels of a quantum particle constrained to move on a ring circunventing a solenoid.
\end{abstract}

\maketitle

\section{Introduction}

Since the establishment of the QED, the effects regarding the the vacuum polarization had been drawing attention mainly in situations with no counterpart in classical Electrodynamics. In this context, we can mention the Uehling potential \cite{Uehling}, which is the electromagnetic potential associated with a single point-like stationary charge corrected in lowest order in the fine structure constant. The calculation of this potential is usually found in the literature perturbatively in the momentum space for a point-like source \cite{GreinerQED,Zee,Kaku,Itzykson} and was also generalized for a charge distribution with finite radius \cite{JPA271994}. Recently, the Uehling potential was calculated exactly in terms of Bessel Integral functions \cite{Frolov}. In the context of non relativistic quantum mechanics, the Uehling potential can lead, for instance, to effects on hydrogen-like atoms \cite{JPB492016}. Atomic effects of QED not accounted by Uehling potential were also studied in the literature \cite{PRA481993}.

Another interesting scenario created by the vacuum polarization, with no counterpart in classical Electrodynamics, is related to the effects which emerge around solenoids and Dirac strings. In this context we can mention, for instance, the radiative corrections to the Aharonov-Bohm scattering \cite{PRD611999}, the interaction between two solenoids \cite{FoundPhys231993}, vacuum currents produced around a Dirac string \cite{PRD471993,IJMP61991}, Bremsstrahlung and pair production in the Aharonov-Bohm potential \cite{IJTP442005} and so on.

The vacuum polarization can be modified by the presence of an external field. The most common situations studied in this context are the effects produced by the presence of an external magnetic field in the Coulomb interaction and in the hydrogen-like atoms \cite{PR1011956,PRA381988,JPB271994,NPB5852000I,NPB5852000II,NPB6092001}. Modifications in the nuclear Coulomb field induced by a strong laser field were also considered in the literature \cite{PRA722005}. 

The vacuum polarization was also studied in coordinate space in reference \cite{PRA892014}, where the Green's function was calculated in higher orders beyond the Uehling term. 

In this paper we study some effects produced by the vacuum polarization of the fermionic field. We focus on situations with no counterpart in classical electrodynamics, studying setups where sources for the electromagnetic field can interact via the vacuum polarization. In section (\ref{Uehlingexato}) we start by studying the standard interaction between two point-like steady charges corrected by the vacuum polarization, in lowest order in the fine structure constant. We calculate exactly the Uehling interaction between them. Our result has two main advantages: it is an exact result (in lowest order in the fine structure constant) valid for any distance between the charges and the result is given by a simple expression written in terms of a K-Bessel functions and Meijer-$G$ functions, what makes it easier to be plotted, once those functions are well known in the literature. 

In section (\ref{campocorda}) we find out the field produced outside a Dirac string due to the vacuum polarization, in lowest order in the fine structure constant. We show that we have a magnetic field outside the string anti-parallel to the internal magnetic flux and we calculate the corrections in order $\alpha$ (the fine structure constant) in the energy levels of a quantum particle constrained to move on a ring (2-D quantum rigid rotor). In section (\ref{cordaatomo}) we show that a hydrogen atom, in its ground state, interact with the string via a kind of Zeeman effect. This interaction falls down very quickly when the distance between the atom and the string increases. This force is attractive when the total angular momentum of the electron is parallel to the internal magnetic flux of the string, and repulsive in the opposite case.

In section (\ref{cordacorda}) we show that it emerges an interaction between two Dirac strings due to the vacuum polarization of the fermionic field. We calculate this interacting force exactly, up to order $\alpha$, for any distance between the strings when they are parallel to each other. Section (\ref{conclusoes}) is devoted to some comments and final remarks.

\section{Uehling interaction}
\label{Uehlingexato}

In this section we calculate exactly the Uehling interaction between two point-like steady charges. 

The gauge sector of the Classical Electrodynamics can be described by the Lagrangian
\begin{equation}
\label{lag}
{\cal L}=-\frac{1}{16\pi}F_{\mu\nu}F^{\mu\nu}-J_{\mu}A^{\mu}-\frac{1}{8\pi}(\partial_{\mu}A^{\mu})^{2}
\end{equation}
where the last term is a gauge fixing one, $A^{\mu}$ is the electromagnetic field, $F^{\mu\nu}=\partial^{\mu}A^{\nu}-\partial^{\nu}A^{\mu}$ is the field strength and $J^{\mu}$ is the external source.

From the Lagrangian (\ref{lag}) one obtain the dynamical equation
\begin{equation}
\label{eqcampoM}
\partial^{\mu}\partial_{\mu}A_{M}^{\nu}=4\pi J^{\nu}
\end{equation}
for which the corresponding propagator $D_{M}(x-y)$ satisfies the differential equation
\begin{equation}
\label{eqpropM}
\partial_{\mu}\partial^{\mu}D_{M}(x-y)=4\pi\delta^{4}(x-y)\ .
\end{equation}
The sub-index $M$ in (\ref{eqpropM}) means that we have the quantities calculated for the Maxwell theory.

The solution for (\ref{eqpropM}) is given by the Fourier integral in the four momentum $p^{\mu}$, 
\begin{equation}
\label{DMFourier}
D_{M}(x-y)=\int\frac{d^{4}p}{(2\pi)^{4}}{\tilde D_{M}}(p)e^{-ip(x-y)}\ ,
\end{equation}
where 
\begin{equation}
\label{defDMp}
{\tilde D_{M}}(p)=-\frac{4\pi}{p^{2}}
\end{equation}
is the Fourier transform of the propagator $D_{M}(x-y)$

The solution for the field equation (\ref{eqcampoM}) is given by the integral
\begin{equation}
\label{solAM}
A^{\mu}_{M}(x)=\int d^{4}x D_{M}(x-y)J^{\mu}(y)\ .
\end{equation}

Substituting the Fourier integrals for the field configuration and for the external source,
\begin{eqnarray}
\label{asd1}
A^{\mu}_{M}(x)=\int\frac{d^{4}p}{(2\pi)^{4}}{\tilde A_{M}}^{\mu}(p)e^{-ipx}\ ,\cr\cr
J^{\mu}(y)=\int\frac{d^{4}p'}{(2\pi)^{4}}{\tilde J}^{\mu}(p')e^{-ip'y}\ ,
\end{eqnarray}
and the integral (\ref{DMFourier}) in Eq. (\ref{solAM}) and using the fact that $\int d^{4}y e^{i(p-p')y}=(2\pi)^{4}\delta^{4}(p-p')$, we can show that
\begin{equation}
\label{zxc1}
{\tilde A_{M}}^{\mu}(p)={\tilde D_{M}}(p){\tilde J}^{\mu}(p)\ .
\end{equation}

With the aid of Eq's (\ref{zxc1}) and (\ref{defDMp}) we can write the Fourier transform of the external source as a function of the Fourier transform of the gauge field obtained from the Maxwell electrodynamics, 
\begin{equation}
{\tilde J}^{\mu}(p)=-\frac{p^{2}}{4\pi}{\tilde A_{M}}^{\mu}(p)\ .
\end{equation}

For a steady external source, $J^{\mu}(x)=J^{\mu}({\bf x})$, one can show that the energy stored in the electromagnetic field is given by
\begin{equation}
E_{M}=\int d^{3}{\bf x}d^{4}y\frac{1}{2}J_{\mu}({\bf x})D_{M}(x-y)J^{\mu}({\bf y})\ .
\end{equation}

It is well known in the literature that the net QED effects of the fermionic vacuum bubbles can be taken into account by a correction in the gauge field propagator \cite{GreinerQED,Itzykson}, as follows
\begin{equation}
\label{asd4}
{\tilde D}(p)={\tilde D_{M}}(p)\big[1+\Pi(p)\big]=-\frac{4\pi}{p^{2}}\big[1+\Pi(p)\big]\ .
\end{equation}
where
\begin{equation}
\label{defPi}
\Pi(p)=-\frac{\alpha}{\pi}\int_{0}^{1}dv\frac{v^{2}\left(1-\frac{1}{3}v^{2}\right)}{v^{2}+\frac{4m^2}{p^2}-1}\ ,
\end{equation}
with $\alpha$ standing for the fine structure constant and $m$, the mass of the electron.

The corrected propagator for the electromagnetic field is given by the Fourier integral
\begin{equation}
\label{asd2}
D(x-y)=\int\frac{d^{4}p}{(2\pi)^{4}}{\tilde D}(p)e^{-ip(x-y)}\ .
\end{equation}

In the presence of an external source, the corrected field configuration is 
\begin{equation}
\label{asd3}
A^{\mu}(x)=\int d^{4}x D(x-y)J^{\mu}(y)\ .
\end{equation}

Substituting the second Eq. (\ref{asd1}) and Eq. (\ref{asd2}) in Eq. (\ref{asd3}), we can show that
\begin{eqnarray}
\label{asd5}
A^{\mu}(x)&=&\int\frac{d^{4}p}{(2\pi)^{4}}{\tilde D}(p){\tilde J}^{\mu}(p)e^{-ipx}\cr\cr
&=&\int\frac{d^{4}p}{(2\pi)^{4}}{\tilde D_{M}}(p)\big[1+\Pi(p)\big]{\tilde J}^{\mu}(p)e^{-ipx}\ .
\end{eqnarray}
where, in the second line, we used Eq. (\ref{asd4}).

With the aid of Eq. (\ref{zxc1}), we can rewrite Eq. (\ref{asd5}) in the form
\begin{eqnarray}
\label{A=AM+DeltaA}
A^{\mu}(x)&=&\int\frac{d^{4}p}{(2\pi)^{4}}{\tilde A_{M}}^{\mu}(p)\big[1+\Pi(p)\big]e^{-ipx}\cr\cr
&=&A^{\mu}_{M}(x)+\int\frac{d^{4}p}{(2\pi)^{4}}{\tilde A_{M}}^{\mu}(p)\Pi(p)e^{-ipx}\ .
\end{eqnarray}

In the second line of Eq. (\ref{A=AM+DeltaA}), the second term on the right hand side can be interpreted as a correction due to the vacuum polarization for the classical field configuration $A^{\mu}_{M}(x)$.

The energy stored in the electromagnetic field due to the presence of a static external source $J^{\mu}({\bf x})$, taking into account the corrections imposed by the fermionic vacuum bubbles, is given by
\begin{equation}
\label{asd6}
E=\int d^{3}{\bf x}d^{4}y\frac{1}{2}J_{\mu}({\bf x})D(x-y)J^{\mu}({\bf y})
\end{equation}

Substituting Eq. (\ref{asd2}) in (\ref{asd6}), using expression (\ref{asd4}), integrating out in $dy^{0}$ and in $dp^{0}$ and using the fact that $\int dy^{0}e^{ip^{0}y^{0}}=2\pi\delta(p^{0})$, we have
\begin{eqnarray}
\label{asd7}
E=E_{M}+\frac{1}{2}\int d^{3}{\bf x}d^{3}{\bf y}\int\frac{d^{3}{\bf p}}{(2\pi)^{3}}J_{\mu}({\bf x})J^{\mu}({\bf y}){\tilde D_{M}}(p^{0}=0,{\bf p})\Pi(p^{0}=0,{\bf p})e^{i{\bf p}({\bf x}-{\bf y})}\ .
\end{eqnarray}

Notice that Eq. (\ref{asd7}) is the energy obtained from the Maxwell electrodynamics, $E_{M}$, with a correction term added. 

Now, let us consider the external source produced by two stationary charges, $q_{1}$ and $q_{2}$, placed at positions ${\bf a}_{1}$ and ${\bf a}_{2}$, respectively,
\begin{equation}
\label{Jduascargas}
J^{\mu}({\bf x})=q_{1}\eta^{\mu0}\delta^{3}({\bf x}-{\bf a}_{1})+q_{2}\eta^{\mu0}\delta^{3}({\bf x}-{\bf a}_{2})\ .
\end{equation}

Substituting Eq's (\ref{Jduascargas}), (\ref{defPi}) and (\ref{defDMp}) in (\ref{asd7}), discarding the terms of self energies (the ones corresponding to the interactions of a given charge with itself) and performing some simple manipulations, we can write
\begin{equation}
\label{zxc4}
E=\frac{q_{1}q_{2}}{a}+4q_{1}q_{2}\alpha\int_{0}^{1}dv v^{2}\Bigl(1-\frac{1}{3}v^{2}\Bigr)\int\frac{d^{3}{\bf p}}{(2\pi)^{3}}\frac{e^{i{\bf p}\cdot{\bf a}}}{{\bf p}^{2}(1-v^{2})+4m^{2}}\ ,
\end{equation}
where we defined the vector ${\bf a}={\bf a}_{1}-{\bf a}_{2}$ and its corresponding modulus $a=|{\bf a}|$.

The right and side of Eq. (\ref{zxc4}) is the same as the one found in the calculations of the Uehling potential \cite{GreinerQED}. We shall calculate it exactly in this section.

The first term on the right hand side of Eq. (\ref{zxc4}) is the coulombian interaction between the charges. This well known result \cite{Zee,BaroneHidalgo1,BaroneHidalgo2,HBB,CBB,BB} is obtained from $E_{M}$ (the interaction energy given by the Maxwell electrodynamics) and the propagator defined in (\ref{defDMp}) and (\ref{DMFourier}). The integral in the second term can be calculated by changing the integration variables ${\bf q}={\bf p}\sqrt{1-v^2}$, as follows
\begin{eqnarray}
\label{zxc3}
\int\frac{d^{3}{\bf p}}{(2\pi)^{3}}\frac{e^{i{\bf p}\cdot{\bf a}}}{{\bf p}^{2}(1-v^{2})+4m^{2}}=\frac{1}{(1-v^{2})^{3/2}}\int\frac{d^{3}{\bf q}}{(2\pi)^{3}}\frac{e^{i{\bf q}\cdot{\bf a}/\sqrt{1-v^2}}}{{\bf q}^{2}+4m^{2}}\ .
\end{eqnarray}
Now we perform the integral in $d^{3}{\bf q}$ with the results of reference \cite{BaroneHidalgo1}. The result is
\begin{equation}
\label{zxc2}
\int\frac{d^{3}{\bf p}}{(2\pi)^{3}}\frac{e^{i{\bf p}\cdot{\bf a}}}{{\bf p}^{2}(1-v^{2})+4m^{2}}=\frac{1}{4\pi}\frac{1}{1-v^{2}}\frac{1}{a}\exp\bigg(-\frac{2ma}{\sqrt{1-v^{2}}}\biggr)
\end{equation}

Substituting the result (\ref{zxc2}) in Eq. (\ref{zxc4}), performing the change of integration variable $u=1/\sqrt{1-v^{2}}$ and making some simple manipulations, we have
\begin{eqnarray}
\label{zxc5}
E&=&\frac{q_{1}q_{2}}{a}+\frac{q_{1}q_{2}}{a}\frac{\alpha}{\pi}\int_{1}^{\infty}du\frac{1}{\sqrt{u^{2}-1}}\Bigl(\frac{2}{3}-\frac{1}{3u^{2}}-\frac{1}{3u^{4}}\Bigr)e^{-2mau}\cr\cr
&=&\frac{q_{1}q_{2}}{a}\Biggl[1+\frac{\alpha}{3\pi}\int_{1}^{\infty}du\frac{1}{\sqrt{u^{2}-1}}\Biggl(2-\frac{1}{u^{2}}-\frac{1}{u^{4}}\Biggr)e^{-2mau}\Biggr]\ .
\end{eqnarray}

All integrals in Eq. (\ref{zxc5}) can be performed exactly. The first one is given by
\begin{equation}
\label{zxc6a}
\int_{1}^{\infty}du\frac{1}{\sqrt{u^{2}-1}}e^{-2mau}=K_{0}(2ma)\ ,
\end{equation}
where $K_{0}(2ma)$ is the K-Bessel function of second kind \cite{Arfken}, and the other two ones are
\begin{eqnarray}
\label{zxc6b}
\int_{1}^{\infty}du\frac{1}{\sqrt{u^{2}-1}}\frac{1}{u^{2}}e^{-2mau}=
\frac{1}{2}(ma)^{3}\mbox{MeijerG}([[], [0]],[[-1/2, -1, -3/2], []],(ma)^{2})\ ,\cr\cr
\int_{1}^{\infty}du\frac{1}{\sqrt{u^{2}-1}}\frac{1}{u^{4}}e^{-2mau}=
\frac{1}{2}(ma)^{5}\mbox{MeijerG}([[], [0]],[[-1/2, -2, -5/2], []],(ma)^2)\ ,
\end{eqnarray}
with $\mbox{MeijerG}$ standing for the G-Meijer functions \cite{Gradshestein}.

Substituting the result (\ref{zxc6a}) in Eq. (\ref{zxc5}) we have finally the Uehling interaction
\begin{eqnarray}
\label{Uehling}
E=\frac{q_{1}q_{2}}{a}\Biggl[1+\frac{\alpha}{3\pi}\Biggr(2K_{0}(2ma)-\frac{1}{2}(ma)^{3}\mbox{MeijerG}([[], [0]],[[-1/2, -1, -3/2], []],(ma)^{2})\cr\cr
-\frac{1}{2}(ma)^{5}\mbox{MeijerG}([[], [0]],[[-1/2, -2, -5/2], []],(ma)^2)\Biggr)\Biggr]\ .
\end{eqnarray}

Notice that expression (\ref{Uehling}) is exact, (up to order $\alpha$) valid for any distance $a$.

It is usual to interpret the result (\ref{Uehling}) as a coulombian interaction between a test charge $q_{2}$ and an effective one given by
\begin{eqnarray}
q_{eff}=q_{1}\biggl[1+g(ma)\biggr]\ .
\end{eqnarray}
where we defined the function
\begin{eqnarray}
\label{defdelta}
g(ma)=\frac{\alpha}{3\pi}\Biggr(2K_{0}(2ma)-\frac{1}{2}(ma)^{3}\mbox{MeijerG}([[], [0]],[[-1/2, -1, -3/2], []],(ma)^{2})\cr\cr
-\frac{1}{2}(ma)^{5}\mbox{MeijerG}([[], [0]],[[-1/2, -2, -5/2], []],(ma)^2)\Biggr)
\end{eqnarray}

\begin{figure}[ht]
   \centering
		\includegraphics[width=0.4\textwidth]{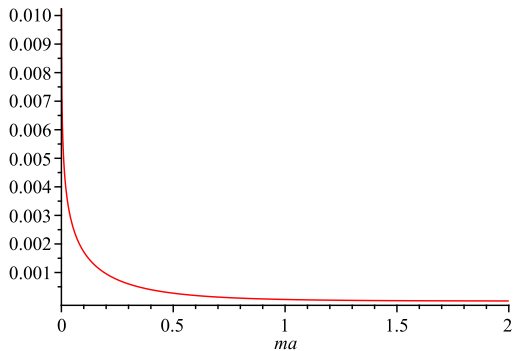}
		\caption{Graphic for the function $g(ma)$, which gives the induced vacuum charge around a point-like electric charge.}
	\label{figdelta}
\end{figure}

In the graphic (\ref{figdelta}) we can see a plot for the function (\ref{defdelta}). It gives the charge distribution induced in the fermionic vacuum by the presence of a point-like stationary electric charge. For $ma\to0$ the function $g(ma)$ diverges as $\ln(ma)$ \cite{Itzykson,GreinerQED}. When $ma\to\infty$ the function $g(ma)$ goes to zero as $e^{-2ma}/(ma)^{3/2}$ \cite{Itzykson,GreinerQED}. 

\section{Field of a Dirac string}
\label{campocorda}

In this section we discuss some effects of the vacuum polarization in the vicinity of a Dirac string. We choose a coordinate system where the Dirac string lies along the $z$-axis with internal magnetic flux $\Phi$. Its corresponding source is given by \cite{BHN,LB1,LB2,Fernanda,Anderson}
\begin{equation}
\label{Dircurr2}
J_{(D)}^{\mu}({\bf x}_{\perp})=\int\frac{d^{4}p}{(2\pi)^{4}}i\Phi(2\pi)^{2}\delta(p^{0})\delta(p^{3})\varepsilon^{0\mu}_{\ \ \nu3}\ p^{\nu}e^{-ipx}\ .
\end{equation}
If $\Phi>0$ we have the internal magnetic field pointing in the $\hat z$ direction. For $\Phi<0$, the internal magnetic field points in the opposite direction.  The sub-index $\perp$ means we are taking just the components of a given vector perpendicular to the string. For instance, ${\bf p}_{\perp}=(p_{x},p_{y},0)$ is the momentum perpendicular to the string.

From Eq. (\ref{Dircurr2}) we can identify the Fourier transform of the Dirac string source
\begin{equation}
\label{defJtilD}
{\tilde J_{(D)}}^{\mu}(p)=i\Phi(2\pi)^{2}\delta(p^{0})\delta(p^{3})\varepsilon^{0\mu}_{\ \ \nu3}\ p^{\nu}\ .
\end{equation}

Substituting the source (\ref{Dircurr2}) in expression (\ref{solAM}), with the Maxwell propagator (\ref{DMFourier}), we have the four-potential 
\begin{equation}
\label{qwe1}
A^{\mu}_{M(D)}(x)=\frac{\Phi}{2\pi}\biggl(0,-\frac{y}{x^{2}+y^{2}},\frac{x}{x^{2}+y^{2}},0\biggr)\ ,
\end{equation}
as expected for a Dirac string. The potential (\ref{qwe1}) produces a vanishing electromagnetic field outside the $z$ axis\footnote{Along the $z$ axis, the magnetic field in infinity.}

The vacuum polarization effects in the vicinity of a Dirac string can be obtained by substituting, in the solution for field configuration (\ref{asd5}), the definitions (\ref{defDMp}) and (\ref{defPi}) and the source for the Dirac string (\ref{defJtilD}). The result is
\begin{equation}
\label{qwe2}
A^{\mu}_{(D)}(x)=A^{\mu}_{M(D)}(x)+\Delta A^{\mu}_{(D)}(x)
\end{equation}
where $A^{\mu}_{M(D)}(x)$ is given by (\ref{qwe1}) and we defined
\begin{eqnarray}
\label{defDeltaAMD}
\Delta A^{\mu}_{(D)}(x)=\int\frac{d^{4}p}{(2\pi)^{4}} \frac{4\pi}{p^{2}}\frac{\alpha}{\pi}\int_{0}^{1}dv\frac{v^{2}\left(1-\frac{1}{3}v^{2}\right)}{v^{2}+\frac{4m^2}{p^2}-1}   
i\Phi(2\pi)^{2}\delta(p^{0})\delta(p^{3})\varepsilon^{0\mu}_{\ \ \nu3}\ p^{\nu}e^{-ipx}\ .
\end{eqnarray}

Integrating out expression (\ref{defDeltaAMD}) in $p^{0}$ and $p^{3}$ and noticing that just its $1$ and $2$ components are non-vanishing, we can write
\begin{eqnarray}
\label{qwe3}
\Delta A^{\mu}_{(D)}({\bf x}_{\perp})&=&4\alpha\Phi\int_{0}^{1}dv\int\frac{d^{2}{\bf p}_{\perp}}{(2\pi)^{2}}-\frac{1}{{\bf p}_{\perp}^{2}}\frac{v^{2}\left(1-\frac{1}{3}v^{2}\right)}{v^{2}-\frac{4m^2}{{\bf p}_{\perp}^2}-1}(0,ip^{2},-ip^{1},0)e^{i{\bf p}_{\perp}\cdot{\bf x}_{\perp}}\cr\cr
&=&4\alpha\Phi\bigg(0,\frac{\partial}{\partial y},-\frac{\partial}{\partial x},0\bigg) \int_{0}^{1}dv\frac{v^{2}\left(1-\frac{1}{3}v^{2}\right)}{1-v^2}\int\frac{d^{2}{\bf p}_{\perp}}{(2\pi)^{2}}\frac{e^{i{\bf p}_{\perp}\cdot{\bf x}_{\perp}}}{{\bf p}_{\perp}^{2}+\frac{4m^2}{1-v^{2}}}\ .
\end{eqnarray}

The integral above is calculated in reference \cite{BaroneHidalgo1}
\begin{equation}
\label{qwe4}
\int\frac{d^{2}{\bf p}_{\perp}}{(2\pi)^{2}}\frac{e^{i{\bf p}_{\perp}\cdot{\bf x}_{\perp}}}{{\bf p}_{\perp}^{2}+\frac{4m^2}{1-v^{2}}}=\frac{1}{2\pi}K_{0}\bigg(\frac{2m|{\bf x}_{\perp}|}{\sqrt{1-v^2}}\bigg)\ ,
\end{equation}
where $K_{0}(x)$ stands for the Bessel function of second kind. So, acting with the derivatives, performing the change of integration variable
\begin{equation}
\label{defxi}
\xi=\frac{1}{\sqrt{1-v^2}}\ ,
\end{equation}
and using the cylindrical coordinates, with $\rho=|{\bf x}_{\perp}|=\sqrt{x^{2}+y^{2}}$, we can write the vector potential (\ref{qwe3}) in the form
\begin{equation}
\label{DeltaAcalculado}
\Delta{\bf A}_{(D)}({\bf x}_{\perp})=
\frac{2^{3}\alpha\Phi m}{3\pi}\int_{1}^{\infty}d\xi\frac{\sqrt{\xi^{2}-1}}{\xi}\bigg(1+\frac{1}{2\xi^{2}}\bigg)K_{1}(2m\xi\rho)\hat\phi\ ,
\end{equation}
with $\hat\phi$ standing for the unitary vector for the azimuthal coordinate.

The potential (\ref{DeltaAcalculado}) is static, has vanishing zero component and does not produce any electric field. Its rotational gives a magnetic field outside the solenoid. It is simpler, first, to calculate the relevant derivatives for the rotational and, after, perform the integration over the $\xi$ variable. The result is,
\begin{equation}
\label{DeltaB}
\Delta{\bf B}=\nabla\times\Delta{\bf A}_{(D)}=-\frac{4\alpha\Phi m^{2}}{3\pi}f(m\rho){\hat z}\ ,
\end{equation}
where we defined the function
\begin{equation}
\label{deff}
f(x)=K_{1}^{2}(x)(1+2x^{2})-2x K_{0}(x)\Big(K_{1}(x)+x K_{0}(x)\Big)\ ,
\end{equation}

\begin{figure}[ht]
   \centering
		\includegraphics[width=0.4\textwidth]{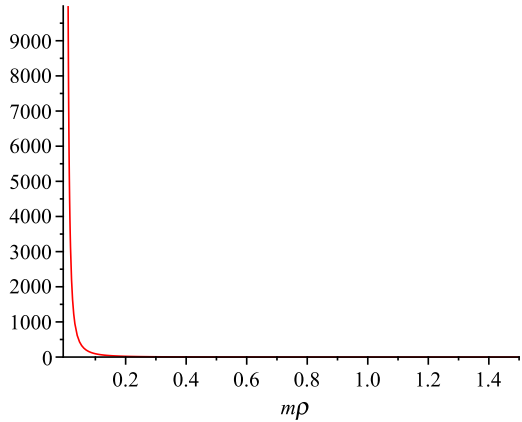}
		\caption{Graphic for the function $f(m\rho)$, which gives the induced magnetic field outside a solenoid.}
	\label{fXx}
\end{figure}

The function (\ref{deff}) is always positive, as one can see in the graphic (\ref{fXx}), so the magnetic field (\ref{DeltaB}) points in the opposite direction in comparison with the magnetic flux of the string.

For small and long distances from the string, the magnetic field (\ref{DeltaB}) reads
\begin{eqnarray}
\Delta{\bf B}\cong-\frac{4\alpha\Phi}{3\pi}\frac{1}{\rho^{2}}{\hat z}\ \ ,\ \ m\rho<<1\cr\cr
\Delta{\bf B}\cong-\alpha\Phi \frac{e^{-2m\rho}}{\rho^{2}}\ \ ,\ \ m\rho>>1 \ .
\end{eqnarray}

\subsection{The 2-D quantum rigid rotor}

The vector potential (\ref{DeltaAcalculado}) induces a modification in the Aharonov-Bohm bound states. It can be seen by taking a very simple example of a 2-dimensional quantum rigid rotor composed by a non-relativistic quantum particle of mass $M$ constrained to move along a circular ring surrounding the Dirac string. Let us take the ring on the plane $z=0$ centered at the origin with radius $b$.

\begin{figure}[ht]
   \centering
		\includegraphics[width=0.4\textwidth]{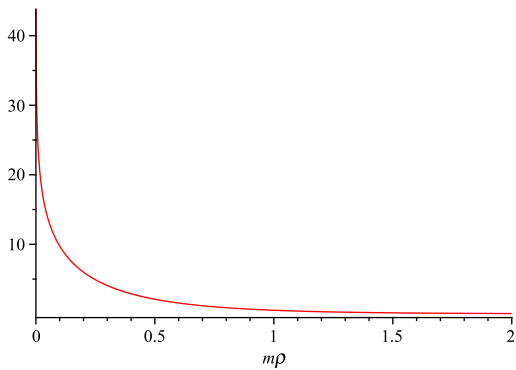}
		\caption{Graphic for the function $F(m\rho)$.}
	\label{FXx}
\end{figure}

As stated in Eq. (\ref{qwe2}), the total vector potential produced by the string is composed by the one obtained from Maxwell theory (\ref{qwe1}) added by the correction (\ref{DeltaAcalculado}). Performing the integral in Eq. (\ref{DeltaAcalculado}), we can write
\begin{equation}
{\bf A}_{(D)}=\frac{\Phi}{2\pi\rho}\Big(1+\frac{4\alpha}{9}F(m\rho)\Big){\hat\phi}\ ,
\end{equation}
where we defined the function
\begin{equation}
\label{defF}
F(x)=(3x^2+4x^{4})K_{0}^{2}(x)-(5x^2+4x^{4})K_{1}^{2}(x)+(6x+4x^{3})K_{0}(x)K_{1}(x) 
\end{equation}
which is always positive, as one can see in the graphic (\ref{FXx}).

It is well known in the literature \cite{GriffithsQM,Sakurai} that the energy levels of a two dimensional quantum rigid rotor are modified when it circumvents an infinite solenoid. In this case we have a very simplified version of the so called Aharonov-Bohm bound states \cite{Sakurai}. In this section we consider the modification introduced by the vacuum polarization in the energy levels of a quantum rigid rotor. For this task, we take a quantum rigid rotor composed by a particle with mass $M$ and electric charge $q$, restricted no move along a ring of radius $b$. We adopt a coordinate system where the ring lies on the plane $z=0$, centered at the origin. We also consider a Dirac string placed along the $z$ axis, with internal magnetic flux $\Phi$. In this case, the Hamiltonian for the charged particle reads
\begin{eqnarray}
\label{defHamiltonianaAB}
H=-\frac{\hbar^{2}}{2Mb^{2}}\frac{d^{2}}{d\phi^{2}}+\frac{i\hbar q\Phi}{2\pi Mb^{2}}\bigg(1+\frac{4\alpha}{9}F(mb)\biggr)\frac{d}{d\phi}+\frac{q^{2}\Phi^{2}}{8\pi^{2}b^{2}}\bigg(1+\frac{4\alpha}{9}F(mb)\biggr)^{2}\ ,
\end{eqnarray}
where $F(mb)$ is defined by (\ref{defF}).

The energy eigenfunctions of the hamiltonian (\ref{defHamiltonianaAB}) are given by
\begin{equation}
\psi(\phi)=Ae^{in\phi}\ ,
\end{equation}
where $n=0,\pm1,\pm2,\cdots$ is any integer and $A$ is a normalization constant. Up to order $\alpha$, the corresponding energy levels are
\begin{equation}
\label{energiasAB}
E=\frac{\hbar^{2}}{2Mb^{2}}\bigg(n-\frac{q\Phi}{2\pi\hbar}\biggr)^{2}-\frac{2\hbar q\Phi\alpha}{9\pi Mb^{2}}F(mb)\bigg(n-\frac{q\Phi}{2\pi\hbar}\biggr)\ .
\end{equation}

The first term on the right hand side of (\ref{energiasAB}) is the well known Aharonov-Bohm energy \cite{GriffithsQM} and the second term is a correction due to the vacuum polarization.

For small and large values of $mb$, the energy (\ref{energiasAB}) reads
\begin{eqnarray}
E\cong\frac{\hbar^{2}}{2Mb^{2}}\bigg(n-\frac{q\Phi}{2\pi\hbar}\biggr)^{2}+\frac{4\hbar q\Phi\alpha}{3\pi Mb^{2}}\ln(mb)\bigg(n-\frac{q\Phi}{2\pi\hbar}\biggr)\ \ ,\ \ mb<<1\cr\cr
E\cong\frac{\hbar^{2}}{2Mb^{2}}\bigg(n-\frac{q\Phi}{2\pi\hbar}\biggr)^{2}-\frac{\hbar q\Phi\alpha}{2 Mmb^{3}}e^{-2mb}\bigg(n-\frac{q\Phi}{2\pi\hbar}\biggr)\ \ ,\ \ mb>>1\ .
\end{eqnarray}

\section{String-atom interaction}
\label{cordaatomo}

The external magnetic field created by a Dirac string can lead to physical phenomena. Let us consider some of its effects produced on a single hydrogen atom. For this task, we take a coordinates system where the atom is placed at the origin and the Dirac string, parallel to the $z$ axis, along the line $(d,0,z)$. In this setup, the Dirac string is placed at a distance $d$ from the atom. We shall restrict to the situation where $d$ is much higher in comparison with the atomic distances. The magnetic field produced in this case can be written by shifting the coordinate $x$ in Eq. (\ref{DeltaB}), as follows 
\begin{eqnarray}
\label{defB'}
\Delta{\bf B}'&=&-\frac{4\alpha\Phi m^{2}}{3\pi}f(m\sqrt{(x-d)^2+y^2}){\hat z}\cr\cr
&\cong&-\frac{4\alpha\Phi m^{2}}{3\pi}f(md){\hat z} \ ,
\end{eqnarray}
where, in the last line, we used the fact that the coordinates $x$ and $y$ are evaluated in the atomic distances and $d>>x,y$.

In a typical experiment $d$ is a macroscopic distance in order of centimeters, what makes the values of the product $md$ very large. So it is legitimate to approximate expression (\ref{defB'}) for $md>>1$, as follows
\begin{eqnarray}
\label{B'aprox}
\Delta{\bf B}'&\cong&-\frac{4\alpha\Phi m^{2}}{3\pi}\frac{3\pi}{4}\frac{e^{-2md}}{(md)^{2}}{\hat z}\cr\cr
&\cong&-\alpha\Phi\frac{e^{-2md}}{d^{2}}{\hat z}\ .
\end{eqnarray}

So, in this setup, the Dirac string produces a kind of Zeeman effect on the atom, once the field (\ref{B'aprox}) is, approximately, constant and uniform along the atom. In this regime, the field produced by the Dirac string is lower than the atomic magnetic field, so we shall use the results of the Zeeman effect (with external magnetic field pointing along the $-{\hat z}$ direction) for weak external field approximation to study the system, where the fine structure constant dominates the energy corrections \cite{Cohen}. We shall take the hydrogen atom on its ground state ($n=1$, $\ell=0$, $m_{\ell}=0$, $j=s=1/2$). In this case, the degeneracy is broken as follows
\begin{equation}
\label{EZeeman}
E=E_{0}\bigg(1+\frac{\alpha^{2}}{4}\bigg)-2m_{j}\mu_{B}|\Delta{\bf B}'|
\end{equation}
where $E_{0}$ is the non-perturbed ground state energy of the hydrogen atom, $\mu_{B}$ is the Bohr magneton and $m_{j}=\pm1/2$ is the azimuthal quantum number for the total angular momentum (for the ground state, $m_{j}=m_{s}$). The minus signal for the Zeeman contribution is due to the fact that the external magnetic field (\ref{B'aprox}) points along the $-{\hat z}$ direction.

As usual, in this case, the contribution of the nuclear magnetic moment (proton) is not relevant \cite{Cohen}. 

Substituting Eq. (\ref{B'aprox}) in (\ref{EZeeman}) we have
\begin{equation}
\label{CDatomo}
E=E_{0}\bigg(1+\frac{\alpha^{2}}{4}\bigg)-2m_{j}\alpha\Phi\mu_{B}\frac{e^{-2md}}{d^{2}}\ .
\end{equation}

The energy (\ref{CDatomo}) exhibits a dependence on the distance $d$ between the Dirac string and the atom and produces a force on the atom (taking the string as fixed) given by
\begin{eqnarray}
\label{FCDatomo}
F&=&-\frac{\partial E}{\partial d}=-4m_{j}\alpha\Phi\mu_{B}\frac{e^{-2md}(md+1)}{d^{3}}\cr\cr
&\cong&-2m_{j}\alpha\Phi\mu_{B}\frac{e^{-2md}}{d^{3}}
\end{eqnarray}
where, in the last line, we discarded a term of order $md$.

When $m_{j}=1/2$, the projection of the total electronic angular momentum of the electron points in the same direction of the internal magnetic flux of the string and the force (\ref{FCDatomo}) becomes negative, what means that it exhibits an attractive nature. When the the total angular electronic momentum points in the opposite direction, with respect to the internal magnetic flux, the force (\ref{FCDatomo}) is positive, what means a repulsive behavior.

In a scattering experiment with a beam of unpolarized atoms propagating in the vicinity of a solenoid, we must have a bifurcation of the beam, according to the total angular momentum of each atom in the beam.

\begin{figure}[ht]
   \centering
		\includegraphics[width=0.4\textwidth]{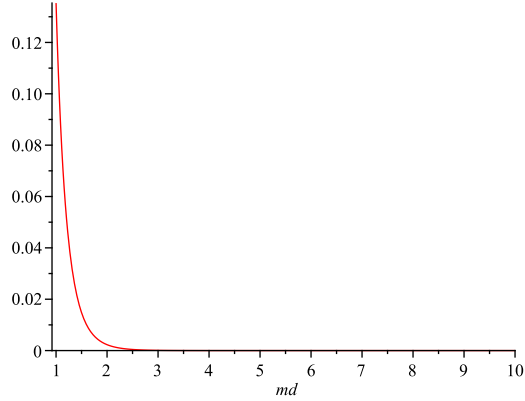}
		\caption{Graphic for the modulus of the force (\ref{FCDatomo}) multiplied by $|2m_{j}\alpha\Phi\mu_{B}m^{3}|^{-1}$ as a function of $md$. Expression (\ref{FCDatomo}) is valid just for large values of $md$.}
	\label{atomo-solenoide}
\end{figure}

From the graphic (\ref{atomo-solenoide}) we can see the behavior of the magnitude of the force (\ref{FCDatomo}) as a function of $md$.  

\section{The interaction between two strings}
\label{cordacorda}
   
In this section we study the interaction between two Dirac strings due to the vacuum polarization. We take two Dirac strings parallel to each other with the first one lying along the $z$ axis and the second one lying along the line $(a^{1},a^{2},z)$. Defining the vector ${\bf a}=(a^{1},a^{2},0)$, we can identify the distance between the strings by the modulus of ${\bf a}$, $\mid{\bf a}\mid=\sqrt{(a^{1})^{2}+(a^{2})^{2}}$. From Eq. (\ref{Dircurr2}), we can write the sources for the two strings, as follows
\begin{eqnarray}
\label{Jduascordas}
J_{(D,1)}^{\mu}({\bf x}_{\perp})=\int\frac{d^{4}k}{(2\pi)^{4}}i\Phi_{1}(2\pi)^{2}\delta(k^{0})\delta(k^{3})\varepsilon^{0\mu}_{\ \ \nu3}\ k^{\nu}e^{-ikx}\cr\cr    
J_{(D,2)}^{\mu}({\bf x}_{\perp})=\int\frac{d^{4}k'}{(2\pi)^{4}}i\Phi_{2}(2\pi)^{2}\delta(k'^{0})\delta(k'^{3})\varepsilon^{0\mu}_{\ \ \nu3}\ k'^{\nu}e^{-ik'x}e^{-i{\bf k'}_{\perp}\cdot{\bf a}}
\end{eqnarray}

Substituting the sources (\ref{Jduascordas}) in Eq. (\ref{asd7}), discarding the self energies of each string, noticing that $E_{M}=0$ (there is no interaction energy between two Dirac strings in the Maxwell Electrodynamics), integrating out in $dy^{0},\ dp^{0},\ dk^{0},\ dk'^{0},\ dk^{3},\ dk'^{3},\ dx^{3}$ and $dp^{3}$, identifying the length of a Dirac string $L=\int dy^{3}$ and performing some simple manipulations, we have the interaction energy between the strings per unit of length
\begin{eqnarray}
\label{qwe5}
{\cal E}&=&\frac{E}{L}=\Phi_{1}\Phi_{2}\int\frac{d^{2}{\bf p}_{\perp}}{(2\pi)^{2}}{\tilde D_{M}}(p^{0}=p^{3}=0,{\bf p}_{\perp})\Pi(p^{0}=p^{3}=0,{\bf p}_{\perp})
\varepsilon^{0\mu}_{\ \ \alpha3}\varepsilon^{0}_{\ \mu\beta3}p^{\alpha}p^{\beta}e^{i{\bf p}_{\perp}\cdot{\bf a}}\cr\cr
&=&-\Phi_{1}\Phi_{2}\int\frac{d^{2}{\bf p}_{\perp}}{(2\pi)^{2}}{\tilde D_{M}}(p^{0}=p^{3}=0,{\bf p}_{\perp})\Pi(p^{0}=p^{3}=0,{\bf p}_{\perp}){\bf p}_{\perp}^{2}e^{i{\bf p}_{\perp}\cdot{\bf a}}\ .
\end{eqnarray}

Now we define the differential operator
\begin{equation}
\nabla_{a}=\bigg(\frac{\partial}{\partial a^{1}},\frac{\partial}{\partial a^{2}},0\bigg)\ ,
\end{equation}
use the fact that ${\bf p}_{\perp}^{2}e^{i{\bf p}_{\perp}\cdot{\bf a}}=-\nabla_{a}^{2}e^{i{\bf p}_{\perp}\cdot{\bf a}}$ and substitute the definitions (\ref{defDMp}) and (\ref{defPi}) in Eq. (\ref{qwe5}), what leads to
\begin{equation}
\label{qwe6}
{\cal E}=4\alpha\Phi_{1}\Phi_{2}\int_{0}^{1}dv\frac{v^{2}\Bigl(1-\frac{1}{3}v^{2}\Bigr)}{1-v^{2}}\nabla_{a}^{2}\int\frac{d^{2}{\bf p}_{\perp}}{(2\pi)^{2}}\frac{e^{i{\bf p}_{\perp}\cdot{\bf a}}}{{\bf p}^{2}_{\perp}+\frac{4m^{2}}{1-v^{2}}}\ .
\end{equation}

Using the result (\ref{qwe4}), acting with the operator $\nabla_{a}^{2}$ and performing the change of integration variables (\ref{defxi}), the energy per unit length (\ref{qwe6}) reads
\begin{eqnarray}
\label{calEduascordas}
{\cal E}=\frac{16}{3\pi}\alpha\Phi_{1}\Phi_{2}m^{2}\int_{1}^{\infty}d\xi\sqrt{\xi^{2}-1}\bigg(1+\frac{1}{2\xi^{2}}\bigg)K_{0}(2m|{\bf a}|\xi)\ .
\end{eqnarray}

The integral in Eq. (\ref{calEduascordas}) can be calculated. The final result for the energy is
\begin{eqnarray}
\label{calEduascordasfinal}
{\cal E}=\frac{4}{3\pi}\alpha\Phi_{1}\Phi_{2}m^{2}\Bigl[K_{1}^{2}(ma)\left(1+(ma)^{2}\right)-2(ma)^{2}K_{0}^{2}(ma)-2(ma)K_{0}(ma)K_{1}(ma)\Bigr]
\end{eqnarray}

The gradient of the energy (\ref{calEduascordasfinal}) with respect to ${\bf a}$ (with a minus signal) gives the force between the strings per unity of length
\begin{equation}
\label{Fduascordas}
{\cal F}=-\nabla_{a}{\cal E}=\frac{8}{3\pi}\alpha\Phi_{1}\Phi_{2}m^{2}\frac{1}{a}\Big[K_{1}^{2}(ma)\Big(1-(ma)^2\Big)+maK_{0}(ma)\Big(maK_{0}(ma)+K_{1}(ma)\Big)\Big]{\hat a}\ ,
\end{equation}
which is repulsive when $\Phi_{1}$ and $\Phi_{2}$ have the same sign and attractive when the magnetic fluxes have opposite signs. In Eq. (\ref{Fduascordas}), ${\hat a}$ stands for the unit vector in the direction of ${\bf a}$ and the function inside brackets is positive along $0\leq ma\leq\infty$.

\begin{figure}[ht]
   \centering
		\includegraphics[width=0.4\textwidth]{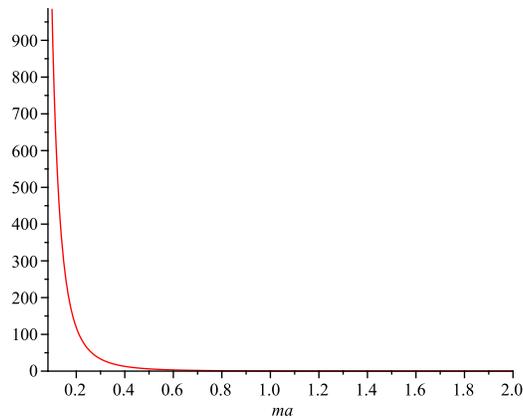}
		\caption{Graphic for the force (\ref{Fduascordas}) divided by $\frac{8}{3\pi}\alpha\Phi_{1}\Phi_{2}m^{3}$, which gives the induced vacuum charge around a point-like electric charge.}
	\label{duascordas}
\end{figure}

Figure (\ref{duascordas}) shows the graphic for the modulus of the force (\ref{Fduascordas}) divided by $\frac{8}{3\pi}\alpha\Phi_{1}\Phi_{2}m^{3}$ as a function of $ma$. For small and large values of $ma$, the behavior of the force (\ref{Fduascordas}) is
\begin{eqnarray}
{\cal F}&\cong&\frac{8}{3\pi}\alpha\Phi_{1}\Phi_{2}\frac{1}{a^{3}}{\hat a}\ \ ,\ \ ma<<1\cr\cr
{\cal F}&\cong&2\alpha\Phi_{1}\Phi_{2}m\frac{e^{-2ma}}{a^{2}}{\hat a}\ \ ,\ \ ma>>1\ .
\end{eqnarray}

As a last comment, we highlight that in some other gauge theories, even without radiative corrections, we can have interactions between Dirac strings and other sources for the gauge field, what is the case of Lee-Wick Electrodynamics \cite{BHN} and theories with explicit Lorentz symmetry breaking \cite{LB1,LB2}.

\section{Conclusions and final remarks}
\label{conclusoes}

In this paper we investigated some effects of the vacuum polarization, in lowest order in the fine structure constant, due to the presence of field sources for the electromagnetic field. All these effects were obtained from the the vacuum polarization tensor of the QED. We obtained an exact expression for the Uehling interaction (Uehling potential) between two point-like steady charges in terms of a K-Bessel function and MeijerG functions. Our results are compatible with that ones obtained in the literature, approximately, for long and small distances between the charges.

We also investigated some phenomena produced outside a Dirac string. One of these effects are the modifications induced in the energy levels of a quantum rigid rotor which circumvents a Dirac string. We have calculated these modifications exactly (for any radius of the quantum rigid rotor) up to order $\alpha$. We have shown that a hydrogen atom interact with a Dirac string via a kind of Zeeman effect. The nature (attractive or repulsive) of this interaction depends on the orientation of the total magnetic moment of the electron with respect to the internal magnetic flux of the string. We restricted to the case where the atom is in its ground state and far away from the string. When the total magnetic moment of the electron is parallel to the internal magnetic flux of the string, the interaction is attractive, on the contrary, it is repulsive. As expected, this interaction is very small and falls down very fast when the distance between the string and the atom increases.

We have also investigated the interaction which emerges between two Dirac strings due to the vacuum polarization. We restricted to the case where the strings are parallel or anti-parallel to each other. We showed that the strings attract each other when their internal magnetic fluxes are anti-parallel. This interaction is repulsive when the internal magnetic fluxes are parallel to each other. We have computed this interacting force exactly (up to order $\alpha$), showing that it falls down vary fast when the distance between the strings increases.

\textbf{\medskip{}
}

\textbf{Acknowledgments.}  F.A. Barone thanks to CNPq (Brazilian agency) under the grant 311514/2015-4 for financial support.



\end{document}